# Evaluating the (in)accessibility of data behind papers in astronomy

Gretchen R. Stahlman
Rutgers University
School of Communication &
Information

## Abstract

This paper presents results of a survey of authors of journal articles published over several decades in astronomy. The study focuses on determining the characteristics and accessibility of *data behind papers*, referring to the spectrum of raw and derived data that would be needed to validate the results of a particular published article as a capsule of scientific knowledge. Curating the data behind papers can arguably lead to new discoveries through reuse. However, as shown through related research and confirmed by the results of the present study, a fully accessible portrait of the data behind papers is often unavailable. These findings have implications for reusability efforts and are presented alongside a discussion of open science.

# Introduction

This paper presents results of a survey of authors of journal articles published over several decades in astronomy. The study focuses on determining the characteristics and accessibility of data behind papers, which refers to the spectrum of raw and derived data that would be needed to validate the results of a particular published article as a capsule of scientific knowledge. Curating the data behind papers alongside paper publication can arguably lead to new discoveries through reuse, especially for fields that require data collected over long timescales such as astronomy and ecology, and for developing robust computational models of changing phenomena over time. However, as shown through prior research and confirmed by the results of the present study, a fully accessible portrait of the data behind papers is often unavailable. The results presented here represent an attempt to measure the accessibility of data behind papers published over time in a particular discipline – astronomy – as a case study and methodological example. The paper will provide a brief literature review to contextualize the motivations and outcomes of this research, followed by methods, analysis, and discussion of results within the framework of open science and reusability, and concluding with suggestions for future research to build upon this effort.

# Background

Encouraging broad accessibility and reuse of research data is a key focus of open science, exemplified by widespread endorsement of the FAIR data principles for making data Findable, Accessible, Interoperable and Reusable (Wilkinson, et al., 2016) and related data publishing initiatives such as the Coalition for Publishing Data in the Earth and Space Sciences (COPDESS) (Hanson, Lehnert & Cutcher-Gershenfeld, 2015). While the capacity of researchers and communities to fully make use of open science resources and practices has been critiqued by scholars such as Ross-Hellauer, et al. (2021), there are demonstrated benefits to sharing datasets, including citation advantages (Piwowar, Day & Fridsma, 2007; Piwowar & Vision, 2013; Henneken & Accomazzi, 2012), further scientific impact through reuse of data (Pasquetto, Randles & Borgman, 2017), transparency and integrity of research (Stockemer, Koehler & Lentz, 2018), and education and training (Geange, et al., 2020). However, it has also been widely demonstrated that data in reusable or reproducible formats often remain largely unavailable to other researchers for legitimate reasons such as concern about misuse or misinterpretation (van Panhuis, et al., 2014; West & Bergstrom, 2021), as well as lack of time (Murillo, 2014), resources (Heidorn, 2008) and career-related incentives (Borgman, 2012, NAS, 2018; Gentemann, et al., 2022).

For the natural and social sciences, data behind papers typically consists of original observational and experimental data generated by researchers themselves or obtained through curated platforms and other researchers, as well as computational simulations and derived data produced through the analysis process. Often data behind papers are provided within the articles through tables, figures and text but are not appropriately packaged, documented or machine readable, requiring extra legwork for someone wishing to reuse or reproduce the work. Data behind papers can range in size and complexity and may fall into the category of "long tail" data that are usually smaller, heterogeneous, and researcher-held data and can often be characterized as niche resources of interest within relatively narrow research areas, and that remain "dark", or in other words inaccessible (Heidorn, 2008; Heidorn, Stahlman & Steffen, 2018). Journal publishers and data curation initiatives increasingly provide mandates, guidelines, and assistance for long-term management of data behind papers, while generic, institutional, and disciplinary repositories are widely available to authors. Organizations including DataCite and Crossref ensure that datasets can receive digital object identifiers (DOIs) to facilitate data citation and linking of data to articles, researchers, and research products in tandem with other services such as ORCID, which assigns unique identifiers to individual authors, and project registration and encapsulation tools such as the Open Science Framework (osf.io) and Whole Tale (wholetale.org).

Amid recent calls to implement policies that overtly enforce research data transparency (West, 2021) – where research products are seen by policy makers to be publicly-funded assets, and where research can have significant societal consequences such as in the biomedical and environmental sciences - Baker & Mayernik (2020) draw a helpful distinction between data production and knowledge production. Within this framework, activities associated with knowledge production (typically realized through communicating research outcomes in journal publications) are often conflated with those of data production where data are intentionally managed and processed to be shared with others. While these two activity streams can overlap significantly, the authors highlight cases of successful knowledge production without focused data production, and data production work that gained importance apart from the primary goal of knowledge production. In other words, the knowledge production stream is historically inherent to the sciences with or without concentrated data production, which throws into question the individual researchers' responsibilities, obligations, and perceptions of making data accessible alongside the needs of the larger research community and the scientific enterprise for transparency, reuse, and efficiency.

# Methodology

To better understand data sharing behavior and openness of data behind papers over time as part of the author's dissertation (Stahlman, 2020), a questionnaire was designed asking corresponding authors of astronomy journal articles published over several decades about the data underlying their specific papers. Following a review and piloting phase, the full survey was released on May 24, 2019, with a completion deadline of June 7, 2019. Qualtrics online survey software was used.

## Sampling

Survey participants were selected from two non-random subsamples. The first subsample includes authors of peer-reviewed papers published in *Publications of the Astronomical Society of the Pacific* (PASP) over approximately a 2-decade period (1994-2016). According to Web of Science, 3,125 papers were published in PASP between 1994 and 2016. Because an initial goal of the project was to perform text classification with a dataset of full-text articles, records corresponding to full-text files that appeared to be incomplete and were removed from the final sample, leaving 2,836 records, of which approximately 1,300 included email addresses for authors and an additional 99 were obtained through manual searching to validate the Web of Science sampling method where email addresses were not present in the bibliographic records.

For the second subsample, related to a parallel research project reported in Stahlman & Heidorn (2020), a Web of Science search for publications associated with National Science Foundation award numbers resulted in 477 peer-reviewed publications under the Astronomy & Astrophysics subject category, and the survey was delivered to corresponding authors of these papers. All papers in this subsample were published between 2016 and 2019, in a variety of astronomy journals.

For the analysis presented here, no distinction was made between the two subsamples of astronomy journal articles. However, it may be worth noting that significant differences between the two groups are age of publication ($t(190.312) = 24.204$, $p = .000$, $n = 211$) and age of respondent ($t(170.366) = 9.230$, $p = .000$, $n = 211$), where the NSF subsample consists of newer papers and younger authors relative to the PASP subsample.

## Questionnaire Design and Implementation

As noted above, Web of Science records include author email addresses in many cases. Piped text was used in the survey introduction to automatically insert a citation to the paper in question, also drawn from Web of Science bibliographic records. In cases where an individual was a corresponding author of several papers, only one paper was selected to ensure that each email address only received one survey request corresponding to one paper. Sixty-four email addresses bounced, and the survey was delivered to unique author email addresses associated with 965 PASP papers. To address the possibility that Web of Science records contain email addresses only for papers with certain characteristics, the survey was also sent to email addresses that were manually retrieved through online searching for 99 additional papers. Including a pilot of 30 recipients, in total the survey was delivered to email addresses associated with 1094 PASP papers.

Separately, to provide insight into the funded research topics and corresponding proposals for the parallel research project (Stahlman & Heidorn, 2020), the same methodology for selecting papers and email addresses was followed, and the survey was successfully delivered to 477 email addresses corresponding to papers funded by NSF grants represented in this sample.

The full survey and descriptive statistics are presently documented in the author's published dissertation (Stahlman, 2020). To summarize the survey instrument, questions were designed with skip logic, and options for responses were provided, overall aiming to understand the nature and whereabouts of underlying data, as well as characteristics and demographic information corresponding to the researchers themselves. The survey was informed by several other surveys of astronomers from which some questions were borrowed with permission (Spuck, 2017; AIP, 2017; AAS, 2013), as well as a series of background interviews with astronomers (Stahlman, 2022), which led to a typology of interconnected data types, formats and observation methods that are referenced in the text of journal articles. Prior to deployment, the survey was reviewed by 7 experts and then piloted with a small group of 30 initial recipients. Question areas most relevant to the present analysis include:

- For this paper, did you utilize new observational data that you collected or collaborated in collecting?
- Did you create simulated data models for this paper?
- For this paper, did you create new or enhanced [derived] data products?

The questionnaire also asked whether each data type is available online for other researchers to access. See the supplementary material for a full list of questions.

## Research Questions and Analysis

Upon the conclusion of the survey period, 211 finished responses were recorded (104 respondents from the PASP subsample and 107 respondents from the NSF subsample). Figure 1 shows the number of responses for each year of paper publication. Survey data was analyzed using SPSS software (version 28) to compile descriptive statistics. The analysis and results presented here were guided by the following research questions:

1. What are the attributes and prevalence of inaccessible data corresponding to publications?

2. How does accessibility of data vary over time since paper publication?

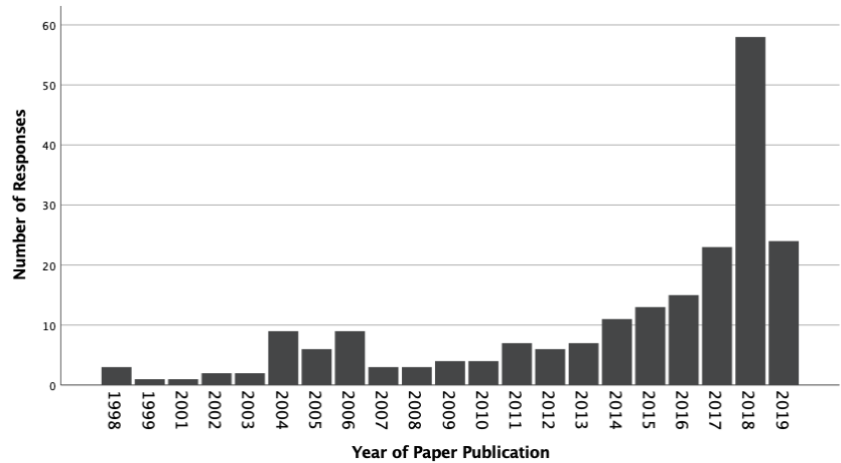

**Figure 1.**   Number of responses received for each publication year.

# Results

While the survey captured variables about data reused by the respondents (via astronomical data archives, data discovered through the literature, and data obtained directly from other researchers), the analysis primarily targets data types for which making the data available is more likely be the responsibility of the researcher(s) as data creators (Pasquetto, et al., 2019). A series of new variables was computed to identify the kinds, formats, and locations of data behind papers generated by the respondents. To better understand the features of papers with inaccessible data, the variables were constructed according to whether a particular paper corresponds to some observational, simulated, and/or derived data available online, or no data in these categories available online. Findings related to the above research questions are presented below.

### RQ1: Attributes and Prevalence of Inaccessible Data

Responses to the following questions indicated high reuse of data:
- For this paper, did you utilize existing data obtained through an astronomical data archive? (53.6%)
- For this paper, did you utilize data that you discovered through a published journal article? (34.6%)
- For this paper, did you utilize data obtained directly from another researcher or team? (36.5%)

Figures 2-4 below show that papers published more recently appear to more frequently demonstrate reuse through the above categories.

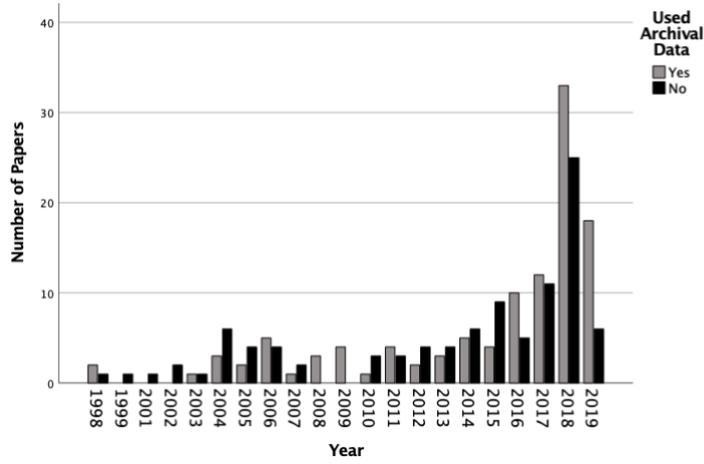

**Figure 2.**         Distribution of papers that utilized archival data for each publication year.

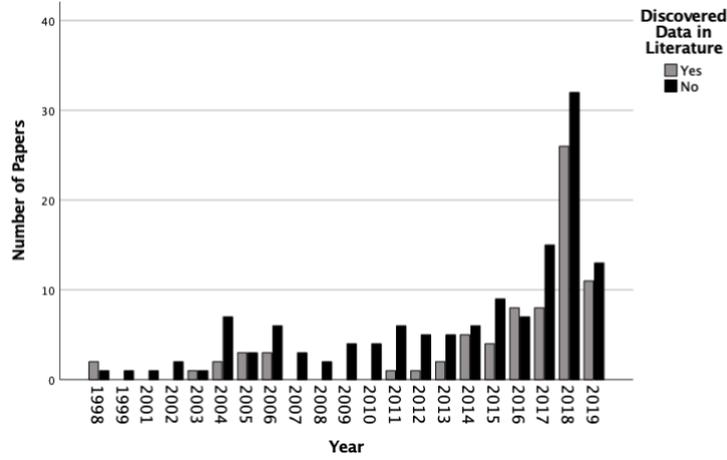

**Figure 3.**         Distribution of papers that utilized data discovered through the literature for each publication year.

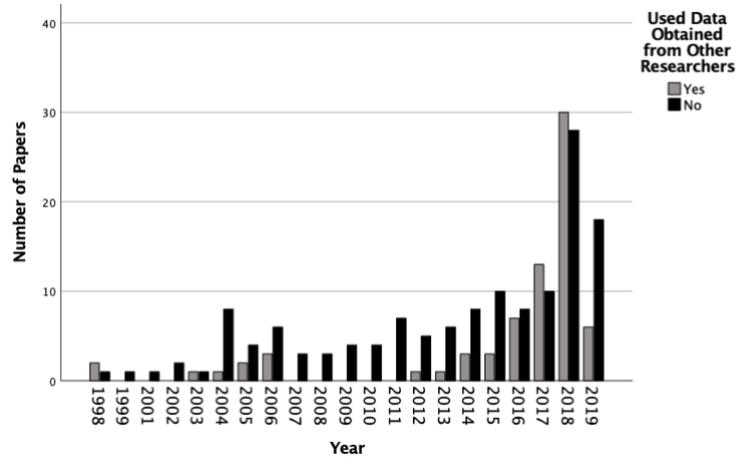

**Figure 4.** Distribution of papers that utilized data obtained from another researcher or team for each publication year.

Accessibility of respondent-generated data was measured through the question areas targeting whether the paper utilized new observational data collected by the author(s), whether the paper utilized simulated data models created by the author(s), and whether new or enhanced data products created for the paper (referred to here as "derived" data). New computed variables show that 69.2% of papers correspond to some data in one or more of the three categories that are not available online for other researchers to access (Table 1). The highest percentage corresponds to the "Simulated Data" category, followed by "Observational Data" and then "Derived Data". The following section will explore differences in these percentages over time since paper publication.

**Table 1.** Accessibility of different categories of data

| Variable | Category | Frequency | Percentage |
|---|---|---|---|
| Observational Data | No inaccessible data | 142 | 67.3 |
| | Some inaccessible data | 69 | 32.7 |
| | N = | 211 | |
| Simulated Data | No inaccessible data | 129 | 61.1 |
| | Some inaccessible data | 82 | 38.9 |
| | N = | 211 | |
| Derived Data | No inaccessible data | 155 | 73.5 |
| | Some inaccessible data | 56 | 26.5 |
| | N = | 211 | |
| Combined (one or more of the above data categories) | No inaccessible data | 65 | 30.8 |
| | Some inaccessible data | 146 | 69.2 |
| | N = | 211 | |

### RQ2: Data Accessibility Over Time Since Publication

The National Science Foundation began requiring Data Management Plans (DMPs) in 2011 (Tenopir, et al., 2011), as other major funding agencies adopted a similar approach. Because this timeframe represents a critical turning point for data management and curation efforts, the data accessibility variables were computed separately for papers published before and after 2011 (Table 2) to get a better sense of trends over time. The frequency of inaccessible data is similar for papers published before 2011 (68.1%) and after 2011 (69.5%), possibly implying that publication date is not directly an indicator of data accessibility. However, the types of inaccessible data do vary over time, with the prevalence of inaccessible observational data higher for older papers and the prevalence of inaccessible simulated and derived data higher for newer papers.

**Table 2.** Data accessibility for papers published before and after 2011

| Data type | Percentage |
|---|---|
| *Before 2011 (n=47)* | |
| Some inaccessible observational data | 40.4% |
| Some inaccessible simulated data | 31.9% |
| Some inaccessible derived data | 14.9% |
| Some inaccessible data (combined) | 68.1% |
| *2011-2019 (n=164)* | |
| Some inaccessible observational data | 30.5% |
| Some inaccessible simulated data | 40.9% |
| Some inaccessible derived data | 29.9% |
| Some inaccessible data (combined) | 69.5% |

### Accessibility Score and "Data Portrait" Case Studies

To further interpret these general findings for the present analysis, a Data Accessibility Score (DAS) was created as a qualitative index for categorizing papers. First, for each data type (a. Observational, b. Simulated, and c. Derived), each paper was assigned a numeric value of 0, 0.5, and 1, depending on whether the paper corresponds to no data available, some data available, or all data available in the category, respectively. Second, the number of data types indicated by the respondent was identified (1, 2, or 3). The DAS ranges from 0 to 1 and was calculated as follows:

$$\frac{\# \: Data \: Types}{a.Obs + b.Sim + c.Der}$$

Figures 5 shows the DAS frequency distribution, where most papers correspond to medium accessibility, followed by no accessible data, and then all data accessible. Figure 6 shows the mean DAS for each year of paper publication, with papers published in 2003 (n=2) having the highest mean DAS and papers published in 2012 (n=12; oddly, just after the DMP implementation timeframe) having the lowest mean DAS. Some missing data resulted from nonresponses to nested questions that were used to compute the data type value, and the total number of observations that were assigned a DAS is 159.

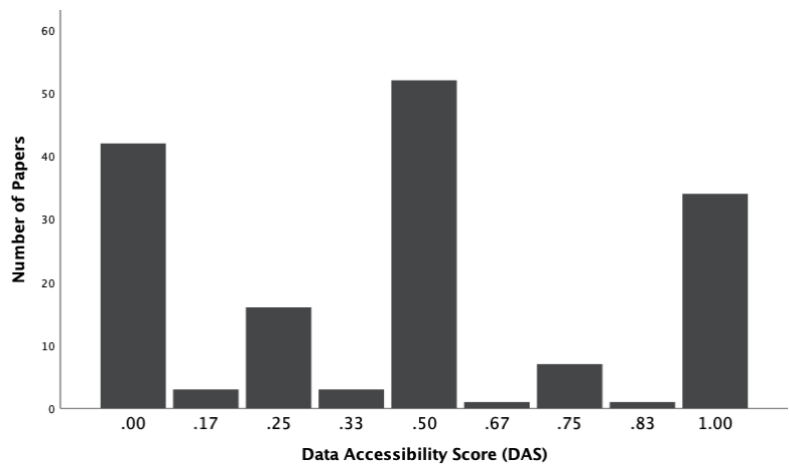

**Figure 5.** Mean Data Accessibility Score (DAS) by number of papers.

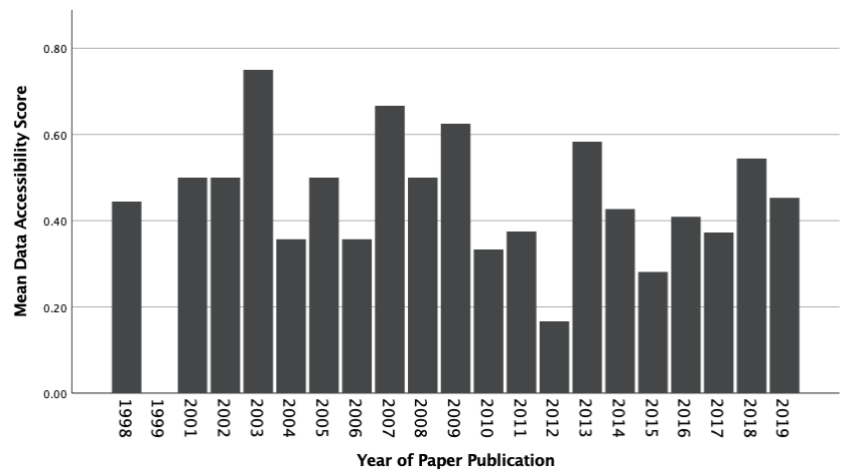

**Figure 6.**     Mean Data Accessibility Score (DAS) for each publication year.

The DAS is illuminating as a first step towards looking more closely at paper trends, but the results alone do not point to a clear pattern of changes in accessibility over time since paper publication and are subject to the self-reporting and question interpretations of the survey respondents. Digging more deeply into unique "portraits" of data behind papers, ten case studies were arbitrarily selected from survey responses with DAS scores and are presented in Tables 3-4, five in each period of interest (before and after 2011) and across the DAS spectrum.

**Table 3.**   Pre-2011 Data Portraits

| 1998-2010 | |
|---|---|
| ID | A |
| DAS | 1 – Fully Accessible |
| Article Category | Research paper or dissertation summary |
| Subject Category | Interstellar Medium and Star Formation |
| Publication Year | 2004 |
| The paper reports conducting new observations to investigate the chemical composition of a nebula. Through the survey, the author indicated that all observational (spectral) data collected are available online, in both an astronomical data archive and institutional repository. The author did not report any other data associated with the paper and indicated that specialized software would not be necessary to reuse the data. While this paper corresponds to fully accessible data according to the author, and the paper itself refers to the two telescope facilities used to collect the data, the paper does not contain links or citations to the indicated repositories or datasets. Both facilities do have online data archives linked through their websites (though neither link appeared to function at the time of this writing on June 9, 2022). The paper acknowledges a software package that was used to reduce the spectroscopic data, which appears to be unsupported. | |
| ID | B |
| DAS | 0.75 – Mostly Accessible |
| Article Category | Astronomical instrumentation, telescopes, observatories and site characterization |
| Subject Category | N/A |
| Publication Year | 2004 |
| The paper reports tests conducted on a type of glass used for near-infrared instrumentation rather than analysis of astronomical research data. Nevertheless, the author's description of the data associated with the paper raise some interesting considerations. Through the survey, the | |

author interpreted the question area inquiring about collecting new observational data to be applicable to the types of observations conducted to measure the refractive index of the glass. The author indicated that the derived/processed data are available online through the published paper in the form of figures and measurements, while the raw data are located offline on an individual or laboratory hard disk. The author also indicated that data were obtained from another researcher in the form of a data catalog. To the question about whether the author would be willing to share data not currently online, the author responded, "No point for a paper with such few citations. The measurements have been superseded in any case". When asked to share any other relevant information about the data, the author further explained, "It is important to distinguish between the raw data, which is not useful to anyone in my opinion, and the processed data, which is published and available in the journal".

| | |
|---|---|
| ID | C |
| DAS | 0.5 – Medium Accessibility |
| Article Category | Research paper or dissertation summary |
| Subject Category | Astronomical software, data analysis, and techniques |
| Publication Year | 1998 |

The paper demonstrates a method of investigating time lags in observations of active galactic nuclei (AGNs). Previous observations of AGNs are analyzed using the new method. Through the survey, the author indicated that archival spectral and light curve data were obtained in FITS (Flexible Image Transport System) and plain text format. The author also indicated that these data were discovered through the astronomical literature and obtained directly from other researchers, which is evident in the paper's citations to previous studies (rather than citations to datasets). The authors created simulated data models, some of which are available online in an astronomical data archive and some of which have been discarded. Specialized software would be required to understand or reuse the data, and the author is "unsure" whether the software is available online. The author would be willing to share any data associated with the paper that are not currently available online.

| | |
|---|---|
| ID | D |
| DAS | 0.25 – Partial Accessibility |
| Article Category | Research paper or dissertation summary |
| Subject Category | Stars and stellar evolution |
| Publication Year | 2004 |

The paper presents observations and measurements about a star with ambiguous spectroscopic classification. The paper refers to the telescope that was used to collect the observations as well as the standard astronomical software used to process the data. Through the survey, the author reported that new observational data were collected for the paper, and none of the data (spectroscopy) are available online for others to access. Rather, the data are stored on an individual or laboratory hard disk and restricted network. Simulated data models were created for the paper, and some of these are available online in an institutional repository, while other simulated data and software are stored on an individual or laboratory hard disk. Specialized software would be required to understand or reuse the data, and the author provided a link to a personal university website where this software is accessible (which is active at the time of this writing on June 9, 2022). The author would be willing to share any data not currently available. The author noted that the sampled paper was an "Interesting choice of paper. I have many other papers that dealt with much larger databases".

| | |
|---|---|
| ID | E |
| DAS | 0 - Inaccessible |
| Article Category | Research paper or dissertation summary |
| Subject Category | Stars and stellar evolution |
| Publication Year | 1998 |

The paper presents observations of a sample of galaxies to determine supernovae rates. Through the survey, the author reported that new observational data were collected for the paper, and none of the data (images) are available online for others to access. Rather, the data are stored on physical media such as plates and tapes. The paper utilized data discovered through the astronomical literature, specified by the author as "Printed figure in a journal article". The paper also utilized data (light curves) obtained directly from another researcher. Simulated data models were created for the paper, which are not online and are stored on an individual or laboratory hard disk. The author would be willing to share data associated with the paper that are not online.

Table 4.   Post-2011 Data Portraits

| 2011-2019 | |
| --- | --- |
| ID | F |
| DAS | 1 – Fully Accessible |
| Article Category | Research paper or dissertation summary |
| Subject Category | Extragalactic Astronomy and Galaxies (Including the Milky Way) |
| Publication Year | 2019 |
| The data-rich paper presents observations of the active galactic nuclei (AGNs) of a sample of galaxies. The observations were conducted using a NASA-operated telescope. Through the survey, the author reported that new observational data were collected for the paper, and some or all data (images and photometry) are available online through an astronomical data archive. Images and spectral data were also obtained and utilized through an astronomical data archive. Additionally, the authors discovered and utilized data (spectral and photometric) through the astronomical literature and obtained data directly from another researcher or team (images, spectral data, and photometric data). Simulated data models were created and are shared on GitHub. The author indicated that other researchers have asked for data associated with the paper, which were shared upon request. Specialized software would be required to understand or reuse the data, which are available online. | |
| ID | G |
| DAS | 0.75 – Mostly Accessible |
| Article Category | Research paper or dissertation summary |
| Subject Category | Stars and stellar evolution |
| Publication Year | 2013 |
| The paper merges new observations of a particular star system with digitized historical data to investigate its spectral properties. Through the survey, the author reported that new observational data were collected for the paper, and some or all of the data (spectroscopy) are available online through an astronomical data archive. Some spectral data are not online and rather are stored on physical media such as plates and tapes. New or enhanced (derived) data products were produced, which are available in an astronomical data archive in FITS format. The author indicated that specialized software would be required to understand or reuse the data but noted that "Reading FITS files is everywhere, but is specialized", meaning that common astronomical data tools are specialized. The author indicated that they would be willing to share data that are not currently available, but noted, "It is effectively impossible to make quality digital scans of photographic astronomical spectra. I'd be delighted to share the problems with anyone who is interested in helping to solve them". | |
| ID | H |
| DAS | 0.5 – Medium Accessibility |
| Article Category | Research paper or dissertation summary |
| Subject Category | Stars and stellar evolution |
| Publication Year | 2011 |
| The paper analyzes observations that were conducted using a NASA-operated telescope to analyze a dwarf nova star. Through the survey, the author reported that new observational data were collected for the paper, and that the data (spectroscopy) are available in a major astronomical data archive to which the author linked. The author also utilized spectral data obtained from the same data archive and obtained directly from another researcher or team. | |
| Simulated data models were created for the paper, with some simulated data and software available via a personal website and some stored on an individual or laboratory hard disk. New or enhanced (derived) data products were created for the paper, but these are not available online. Rather, the data (images and computational model(s) or simulation(s)) are stored on an individual or laboratory hard disk. Specialized software would be needed to understand or reuse the data, and the author provided a link to a common suite of tools commonly used by astronomers. The author would be willing to share any data not currently available, but noted, "Most of the data is available online and I don't think many people are really interested in that data". | |
| ID | I |
| DAS | 0.25 – Partial Accessibility |
| Article Category | Research paper or dissertation summary |
| Subject Category | Astronomical software, data analysis, and techniques |
| Publication Year | 2013 |
| The paper presents a detailed analysis of an astrometry method. Through the survey, the author reported that new observational data were collected for the paper, and that the data (images) are available in a major astronomical data archive. Other observational data (photometry) are not available online, and the author is "unsure" of the location. The author also indicated that archival image data from the "ESO [European Southern Observatory] Database" were utilized for the paper. Simulated data models were created for the paper, which are not available online and are located on an individual or laboratory hard disk. The author indicated that specialized software would be required to understand or reuse the data and the software is not available online. The author would be willing to share the software as well as any data associated with the paper that are not online. The author explains, "I used a reference that is not publicly available, part of the restricted [observatory name redacted] technical archive (considered an industrial secret). The paper was confidentially sent to me by a colleague". | |
| ID | J |
| DAS | 0 - Inaccessible |
| Article Category | Research paper or dissertation summary |
| Subject Category | Astronomical software, data analysis, and techniques |
| Publication Year | 2012 |
| The paper demonstrates a method for instrument calibration. Through the survey, the author reported that new observational data were collected for the paper, and that the data (images and spectral data) are not available online. Rather, the data are stored on physical media such as tapes and plates and on an individual or laboratory hard disk. The author would be willing to share the data, but notes, "This was a 'techniques' paper and the data associated with it is simple and easily reproducible. Unlike many papers, which I and others write, they can contain lots of direct observational data that is not easy to reproduce". | |

## Discussion and Conclusion

This study has demonstrated a method to characterize the accessibility of different types of data behind papers by asking authors about the research processes that produced specific journal articles, interpreted alongside the articles themselves. The general findings suggest that a large portion of papers in astronomy corresponds to some inaccessible data, and that the prevalence of inaccessible data is consistent for newer and older papers while types of inaccessible data vary. For example, the survey illustrates that newer papers in particular may correspond to inaccessible simulated data models, which aligns with recent conversations about the importance of FAIR-izing computationally intensive models (Mullendore, et al., 2021; Alvarez, et al., 2022). The survey also shows evidence that older papers may correspond more frequently to inaccessible observational data, which aligns with calls for legacy data rescue (Steffen & Hunt, 2022; Griffin, 2015). The case study exploration presented here further provides a focused lens on survey responses, highlighting a series of "data portrait" scenarios over time since paper publication. The case study portraits illustrate a muddled reality of data behind papers, where a variety of products and processes converge to produce a paper with a mixture of accessibility. A key theme that emerged is that authors have a strong sense of what should be shared and why at any given point in time. The authors of papers "B" and "H" did not see a point to publishing data because of low attention to the papers and data, with the author of paper "B" noting that the most useful processed data are available through tables and figures in the published article, while the author of paper "I" noted that the data associated with the paper is "simple and easily reproducible". Another key theme is that curation of data behind papers is a moving target as technology evolves. Paper "A" received a DAS of "1 – Fully Accessible" based on the author's report, but the data are no longer accessible due to non-functioning websites and unsupported software. Regarding format migration for older data, the author of paper "G" pointed out critical quality issues with digitized historical data. Paper "F", recently published, represents the current ideal of open science by utilizing well-curated data archives and sharing software on GitHub. However, as demonstrated by paper "A", ongoing accessibility over time will depend on the long-term sustainability of these platforms.

Linking journal articles to associated datasets and software has increasingly become a priority and a challenge for publishers, funders, researchers, and other stakeholders within the evolving ecosystems of research policy and contemporary scholarly communication. From a policy perspective, assessing and justifying the scholarly impact of published research outcomes requires availability of research products including data. From a curation perspective, data are information resources and opportunities for discovery and reuse through long-term stewardship. From an institutional perspective, data represent assets to be protected, managed, and leveraged strategically. And from an epistemological perspective, data serve as indicators of research quality and dynamic building blocks for further investigations. These complementary perspectives and agendas inherently converge through open science tools, infrastructures, and practices that aim to facilitate more widespread data sharing and linking than was previously possible or feasible (Fecher & Friesike, 2014). Nevertheless, the survey responses appear to reflect a persistent focus on knowledge production over thorough data production (Baker & Mayernik, 2020). This suggests that open science policies and practices may need to continue to accommodate the integral role of "trust" in scholarly communication (Pasquetto, et al., 2019) and potentially inevitable distinctions between credible evidence, reusable data, and reproducible research.

The presented results are entwined with the unique knowledge infrastructures of astronomy (Borgman & Wofford, 2021). However, the study methodology could be adapted in other disciplines and to inform curation activities more broadly by interfacing directly with the data creators through their published papers. The study also has some key limitations. First, the relatively small size of the survey (n=211), non-random sampling, and nested question areas with few responses limit statistical analysis. At times, respondents seemed confused about what was meant by questions targeting "observational", "new/enhanced" and "simulation" data as well as "specialized software". Future work may leverage these lessons learned to conduct a similar survey across disciplines for a more comprehensive understanding of the accessibility and curation needs of various types and formats of data behind papers. Future work will also further explore the survey through additional analysis and beyond, including investigating the topic of usefulness of various data products in astronomy and other communities to inform open science efforts. Finally, the anonymized survey itself should be documented and published as a data product for reuse following sufficient dissemination of the additional results.

# References

AIP Statistical Research Center (2017). *The Longitudinal Study of Astronomy Graduate Students.* https://www.aip.org/statistics/lsags

Alvarez, M., Bailey, S., Bard, D., Gerhardt, L., Guy, J., Juneau, S., ... & Weaver, B. (2022). Data Preservation for Cosmology. *arXiv preprint arXiv:2203.08113.*

American Astronomical Society (2013). *AAS Biennial Demographics Survey.* https://aas.org/posts/news/2014/01/2013-aas-biennial-demographics-survey-released

Baker, K. S., & Mayernik, M. S. (2020). Disentangling knowledge production and data production. *Ecosphere, 11*(7), e03191.

Borgman, C.L. (2012). The conundrum of sharing research data. *Journal of the American Society for Information Science and Technology 63*(6).

Borgman, C. L., & Wofford, M. F. (2021). From data processes to data products: Knowledge infrastructures in astronomy. *Harvard Data Science Review, 3*(3). https://doi.org/10.1162/99608f92.4e792052

Daston, L. (1995). The moral economy of science. *Osiris, 10*(1), 2-24. doi:10.1086/368740

Fecher, B., & Friesike, S. (2014). Open science: one term, five schools of thought. *Opening science*, 17-47.

Geange, S. R., von Oppen, J., Strydom, T., Boakye, M., Gauthier, T. L. J., Gya, R., ... & Vandvik, V. (2021). Next-generation field courses: Integrating Open Science and online learning. *Ecology and Evolution, 11*(8), 3577-3587.

Gentemann, C., Erdmann, C., & Kroeger, C. (2022). Opening Up to Open Science. *Issues in Science and Technology, 38*(3).

Griffin, R. E. (2015). When are old data new data?. *GeoResJ, 6*, 92-97.

Hanson, B., Lehnert, K., & Cutcher-Gershenfeld, J. (2015). Committing to publishing data in the Earth and space sciences. *Eos, 96*(10.1029).

Heidorn, P.B. (2008). Shedding light on the dark data in the long tail of science. *Library Trends 57*(2), 280-299.

Heidorn, P. B., Stahlman, G. R., & Steffen, J. (2018). Astrolabe: curating, linking, and computing astronomy's dark data. *The Astrophysical Journal Supplement Series, 236*(1), 3.

Henneken, E. & Accomazzi, A. (2012). Linking to data: Effect on citation rates in astronomy. *ASP Conference Series, 461*, 763-766.

McCray, W. (2014). How astronomers digitized the sky. *Technology and Culture, 55*(4), 908-908. doi:10.1353/tech.2014.0102

Mullendore, G. L., Mayernik, M. S., & Schuster, D. C. Open Science Expectations for Simulation-Based Research. *Frontiers in Climate*, 162.

Murillo, A.P. (2014). Data At Risk Initiative: Examining and facilitating the scientific process in relation to endangered data. *Data Science Journal, 12*, 207-219.

National Academies of Sciences, Engineering, and Medicine. (2018). *Open science by design: Realizing a vision for 21st century research.* National Academies Press.

Pasquetto, I. V., Randles, B. M., & Borgman, C. L. (2017). On the reuse of scientific data. *Data Science Journal, 16*, 8.

Piwowar, H.A. & Vision, T.J. (2013). Data reuse and the open data citation advantage. *PeerJ, 1*, e175.

Piwowar, H.A., Day, X., & Fridsma, X. (2007). Sharing detailed research data is associated with increased citation rate. *PloS one, 2*(3), e308.

Ross-Hellauer, T., Reichmann, S., Cole, N. L., Fessl, A., Klebel, T., & Pontika, N. (2021, July 8). Dynamics of Cumulative Advantage and Threats to Equity in Open Science - A Scoping Review. https://doi.org/10.31235/osf.io/d5fz7

Spuck, T. S. (2017). *What do astronomers do: A survey of U.S. astronomers' attitudes, tools and techniques, and social interactions engaged in through their practice of science* (Order No. 10616586). Available from ProQuest Dissertations & Theses Global. (1948887650).

Stahlman, G. R. (2022). From nostalgia to knowledge: Considering the personal dimensions of data lifecycles. *Journal of the Association for Information Science and Technology.* https://doi.org/10.1002/asi.24687

Stahlman, G. R. (2020). *Exploring the long tail of astronomy: A mixed-methods approach to searching for dark data* (Doctoral dissertation, The University of Arizona).

Stahlman, G. R., & Heidorn, P. B. (2020). Mapping the "long tail" of research funding: A topic analysis of NSF grant proposals in the division of astronomical sciences. *Proceedings of the Association for Information Science and Technology, 57*(1), e276.

Steffen, J., & Hunt, S. E. (2022). Tapes and Papers: An Astronomical Heritage Archival Project of the American Astronomical Society and NSF's National Optical-Infrared Astronomy Research Laboratory. *Bulletin of the American Astronomical Society, 54*(2).

Stockemer, D., Koehler, S., & Lentz, T. (2018). Data Access, Transparency, and Replication: New Insights from the Political Behavior Literature. *PS: Political Science & Politics, 51*(4), 799-803.

Tenopir, C., Allard, S., Douglass, K., Aydinoglu, A. U., Wu, L., Read, E., ... & Frame, M. (2011). Data sharing by scientists: practices and perceptions. *PloS one, 6*(6), e21101.

Van Panhuis, W. G., Paul, P., Emerson, C., Grefenstette, J., Wilder, R., Herbst, A. J., ... & Burke, D. S. (2014). A systematic review of barriers to data sharing in public health. *BMC public health, 14*(1), 1-9.

West, J. (2021). *Cautionary notes for data-driven science policy.* Retrieved from: https://jevinwest.org/publications.html

West, J. D., & Bergstrom, C. T. (2021). Misinformation in and about science. *Proceedings of the National Academy of Sciences, 118*(15).

Wilkinson, M. D., Dumontier, M., Aalbersberg, I. J., Appleton, G., Axton, M., Baak, A., ... & Mons, B. (2016). The FAIR Guiding Principles for scientific data management and stewardship. *Scientific data, 3*(1), 1-9.